\documentclass[letter,english]{article}
\usepackage{emulateapj,onecolfloat,psfig}

\def\jcoph{J. Comp.\ Phys.}

\def\spose#1{\hbox to 0pt{#1\hss}} 
\def\ltsim{\mathrel{\spose{\lower.5ex\hbox{$\mathchar"218$}}
     \raise.4ex\hbox{$\mathchar"13C$}}}
\def\gtsim{\mathrel{\spose{\lower.5ex \hbox{$\mathchar"218$}}
     \raise.4ex\hbox{$\mathchar"13E$}}}

\def\eg{{\it e.g.}}
\def\etal{{\it et al.}}

\def\ie{{\it i.e.}}
\slugcomment{Revised version to appear in ApJ Letters}

\begin{document}

\twocolumn
[
\title{Long-lived lop-sided modes of annular disks orbiting a central mass}
\author{V. Jacobs and J. A. Sellwood}
\affil{Rutgers University, Department of Physics \& Astronomy, \\
       136 Frelinghuysen Road, Piscataway, NJ 08854-8019 \\
       {\it vjacobs/sellwood@physics.rutgers.edu}}

\begin{abstract}
We demonstrate the existence of long-lived, large-amplitude eccentric modes in massive, annular particle disks in orbit about a central mass.  The lopsided modes we have found precess slowly in the prograde direction at a rate which increases with disk mass and decreases with the amplitude of the distortion.  The lopsidedness generally survives for as long as we run the calculations and may last indefinitely; we found no significant decay in over 700 disk-particle orbits in one case.  These strongly non-linear modes are readily created using a number of different perturbing rules, but we find no evidence for linear instabilities in unperturbed disks.  Our results suggest that Tremaine's eccentric disk model for the nucleus of M31 may be viable, but none of our models matches all aspects of the observational data.
\end{abstract}

 ]

\section{Introduction}
Massive compact objects are frequently surrounded by disks; familiar examples are in accretion disks, protoplanetary disks and planetary ring systems.  The disk generally has a small mass in comparison with the central object, but in some cases it may have sufficient mass to cause a significant departure from a Keplerian rotation law.  The possible small-amplitude, lop-sided ($m=1$) modes of such disks have been studied by Adams, Ruden \& Shu (1989), Shu \etal\ (1990), Tremaine (2000) and others, while Lee \& Goodman (1999) report a study of possible finite-amplitude, one-armed spiral modes.  Ogilvie (2001) presents a general treatment of non-linear modes in fluid disks.

Large-amplitude eccentric modes in near-Keplerian particle disks have received special attention since Tremaine (1995) proposed an eccentric disk model for the nuclear structure of the Andromeda galaxy (M31).  Sridhar, Syer \& Touma (1999) and by Salow \& Statler (2000) attempt to construct self-consistent models, while Bacon \etal\ (2000) have conducted $N$-body simulations.

We here report our own $N$-body simulations of massive, eccentric disks in orbit about a central point-like mass.  We find it surprisingly easy to create long-lived, large-amplitude eccentric modes in annular disks having a wide range of masses (up to $\sim 50\%$ of the central mass), radial extents, and eccentricity profiles.  Since low-mass disks are expected to be cool and thin, we confine the particles in our simulations to motion a plane; higher mass disks, which are likely to possess larger amounts of random motion, may have a non-negligible thickness and require a 3-D code.  In the final section, we make a preliminary comparison between our simulations and the data obtained by Kormendy \& Bender (1999) on the M31 nucleus.

\section{Initial conditions and methods}
Our simulations are of an isolated system with two mass components: a razor-thin disk of particles whose motion is confined to a plane and a mobile central mass.  The central object is a slightly softened point mass, $M_\bullet$, having the Plummer density profile
\begin{equation}
\rho_{\bullet}(r) = M_\bullet {3b^2 \over 4\pi(r^2 + b^2)^{5/2}},
\end{equation}
with a small core radius $b$.   
As our initial experiments with filled disks did not produce any long-lived modes, we here report simulations with annular disks only.  Our unperturbed disk has the surface density distribution
\begin{equation}
\Sigma(R) = {M_d \over \pi^2 a^2 R_0} \left[a^2 - \left( R - R_0 \right)^2 
\right]^{1/2}
\end{equation}
for $|R-R_0|<a$ and is zero outside this range.  Before we introduce eccentricity, the orbits of all particles are circular.  We adopt $M_\bullet$ as our mass unit and $R_0$ as our length unit and choose $b = 5 \times 10^{-3}R_0$.  We set $G=1$, so that our time unit is $(R_0^3/GM_\bullet)^{1/2}$.  The orbital period in the middle of the annulus is therefore $\sim 2\pi$.

In order to facilitate the creation of an initially eccentric disk, we begin all our simulations with $M_d/M_\bullet = 10^{-3}$, and then increase the disk mass to the desired value during the first part of the evolution.

We make a low-mass disk eccentric by adjusting the orbit of each disk particle to be a Kepler ellipse aligned with the $x$-axis and with the central mass at one focus -- \ie\ we neglect the disk contribution to the total potential.  The particles of the annular disk (eq.\ 2) are moved and given new velocities such that their elliptical orbits intercept the positive $x$-axis where their original circular orbit would have.  Our standard rule (a), sets the distribution of eccentricities such that all initial orbits pass through the point $(x,y)=-(R_0 + a, 0)$ -- \ie\ the outer edge of the disk remains circular while the elliptical inner edge has semi-major axis $R_0$ and eccentricity $e = a/R_0$.  We have experimented with other rules: (b) making all orbits pass through a point outside the original disk so that even the outer edge is eccentric, (c) a fixed fraction of the eccentricity required by these rules, (d) a disk that is eccentric only in the inner parts and axisymmetric farther out, (e) a constant eccentricity throughout the disk, and (f) a reversed trend with radius so that the inner edge is circular.

We allow the low-mass disk to evolve for a few orbital periods, although little happens during this time.  We then raise the mass of each particle so that the disk mass rises at a constant rate.  Generally, this rate is $dM_d / dt = 10^{-3}M_\bullet$, so that it takes $\sim 16$ orbital periods for the disk mass to rise by $0.1M_\bullet$; other rates were also tried.  Once the desired mass is reached, we reposition our coordinate origin at the center of mass and subtract any net momentum from the system.  The subsequent evolution is unrestrained.

We evaluate gravitational forces between the disk particles using a grid method for efficiency (Sellwood 1997).  A 2-D Cartesian grid (\eg\ Hohl \& Hockney 1969) is ideal for this case because the disk is thin and not centrally concentrated, and the grid has no preferred center.  We determine the forces between the central mass and disk particles directly and compute the motion of the central mass in response to the total force from the disk.  We sub-divide the time steps of disk particles as they approach the central object.  The code conserves energy and both linear and angular momentum to a high degree of precision.  We employ $100\,000$ disk particles, a $256^2$ grid, and a time step of 0.01 or 0.005.  We have verified that our results are not significantly different when the grid size, particle number or time step is varied by factors of two or more.

\begin{figure}[t]
\psfig{figure=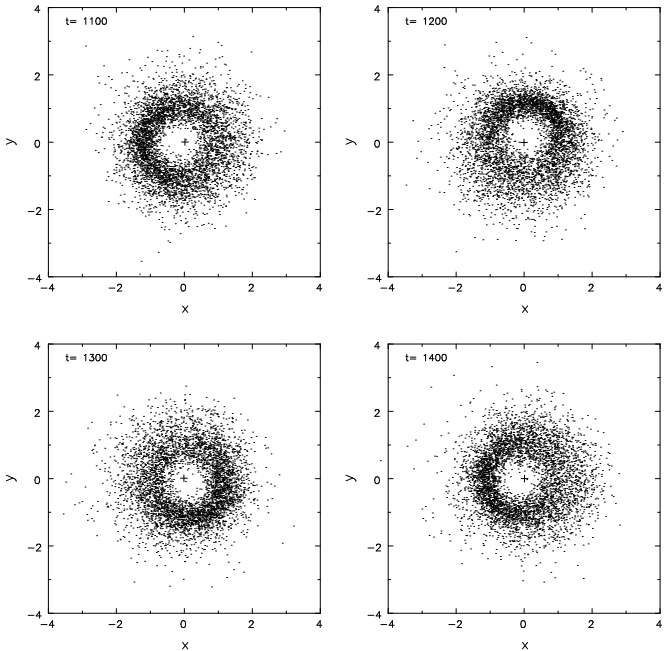,width=\hsize,angle=0,clip=}
\caption{\small A short sequence of snapshots from our canonical model (only 1 particle in 20 is plotted in each frame).  The position of the central mass is marked by the plus symbol.}
\label{fig:image}
\end{figure}

\section{Results}
Our canonical model has a final disk mass of $0.1M_\bullet$, $a=R_0/2$ and the initial eccentricity prescription (a) above (\ie\ the ratio of apocenter to pericenter radius is three for the innermost orbit).  As the disk mass was increased, the initially neat elliptical stream pattern developed transient spiral patterns, heated somewhat and spread radially.  As soon as the disk mass stopped increasing (at $t=200$), the model settled to a lop-sided disk with no significant spirality, as shown in Figure \ref{fig:image}.

The major axis of the eccentric disk precessed at an almost steady rate with no obvious decay.  To quantify this behavior, we form the Fourier coefficients
\begin{equation}
A(m,\gamma,t) = {1 \over N} \sum_{j=1}^N \exp\left\{ im\left[\phi_j + \tan\gamma \ln \left( {R_j \over R_0} \right) \right] \right\},
\end{equation}
where $(R_j,\phi_j)$ are polar coordinates of the $j$-th particle at time $t$.  The amplitude of the $m=1$ distortion decreased slightly from its initial value as the disk relaxed while the mass was increased.  Subsequently, the amplitudes of the $m=1$ coefficients manifest a mild modulation, suggesting beats between two waves with a gradually increasing beat period.  The frequencies (pattern speeds) of the stronger and weaker waves are $\Omega_p \sim 0.045$ and $\Omega_p \sim 0.008$ in our units, in the first part of the evolution, but the stronger wave later slows to $\Omega_p \sim 0.040$.  The amplitude of neither wave decays significantly during the last $\sim 4500$ dynamical times -- these disturbances appear to be indefinitely-lived, large-amplitude, lop-sided modes.

\begin{figure}[t]
\centerline{\psfig{figure=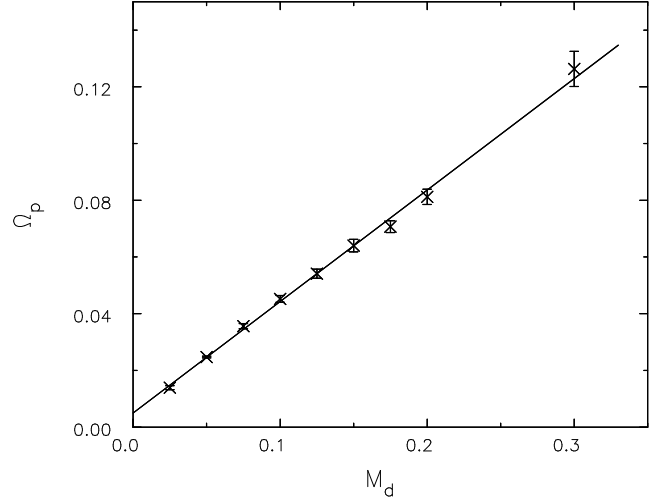,width=\hsize,angle=0,clip=}}
\caption{Variation of pattern speed with disk to central mass ratio.  The best linear fit is shown by the line.}
\label{fig:pattspd}
\end{figure}

We were able to find long-lived modes in most of the annular-disk models we tried.  Models with $a=0.5R_0$ and a higher disk mass behaved reasonably well for $M_d \ltsim 0.3M_\bullet$.  The large random motions in disks of yet higher mass cause small numbers of particles to become unbound; the surviving disk appears to remain lop-sided, but the drift in the center of mass of the bound particles complicates the analysis.  The pattern speed of the dominant mode, which is always the fastest, increases in an approximately linear manner with the disk mass (Figure \ref{fig:pattspd}), behavior that was predicted by Tremaine (2000) for small-amplitude waves.  A linear fit yields $\Omega_p \simeq (0.393 \pm 0.008) M_d/M_\bullet$.  We stress that our results are for waves of considerable amplitude and we report below that less strongly perturbed disks have higher pattern speeds, in agreement with Salow \& Statler (2000).

Reducing the initial eccentricity, but holding $a=R_0/2$ and $M_d/M_\bullet = 0.1$, increased the pattern speed of the resulting mode but had little effect on its survivability, at least until the distortion was very mild.  A lopsided mode stood out clearly in a model in which the initial eccentricty of all orbits was reduced to 20\% of that required for rule (a), but a long-lived mode was barely detectable when the initial distortion was reduced to 10\%.

In fact, we were able to generate long-lived eccentric disks with quite remarkable ease.  Lopsidedness survived, but usually with different pattern speeds, in all disks having strong initial eccentricities, for almost all types of initial perturbation.  We also varied the radial extent of the disk in the range $1/2 \leq a/R_0 \leq 7/9$; low-mass disks settled to a long-lived mode in every case.  Cases in which the lop-sidedness weakened considerably were for more massive disks ($\gtsim 0.3M_\bullet$) with larger radial extents ($a/R_0 \gtsim 2/3$), whether the initial perturbation included all or just part of the disk.

Statler (1999) defines orbital eccentricity $e = (x_-+x_+)/(x_--x_+)$, where $x_+$ and $x_-$ are respectively the intercepts of an aligned periodic orbit on the positive and negative $x$-axis.  He argues that a self-consistent mode requires $e$ to change sign near the densest part.  We use the positions and velocities of the particles at one instant to define Kepler ellipses for each.  As they are, in general, misaligned with the principal axis of the mode, we define $e$ from the intercepts with the ellipse major axis.  The disk appears to have two populations of particles, a strongly eccentric group which all have $e \gtsim 0.7$ and are generally aligned with the mode, and a ``hot'' population of particles with $-0.4 \ltsim e \ltsim 0.4$ whose orbits are aligned randomly with respect to the major axis of the mode.  This structure appears to be generic, although the relative proportions of the two populations in different models vary.  Thus, we find no evidence of a coherent change in orientation in the dense part of the disk.

We were unable to detect any growing modes in a disk with no initial distortion; transient spiral patterns as the disk mass increased heated the particle distribution to $Q \sim 3$ in the densest part of the disk, but it then remained closely axisymmetric.  Tremaine (2000) predicts that small-amplitude modes should be stable, and our results also suggest that no lopsided instability is present.

\begin{figure}[t]
\centerline{\psfig{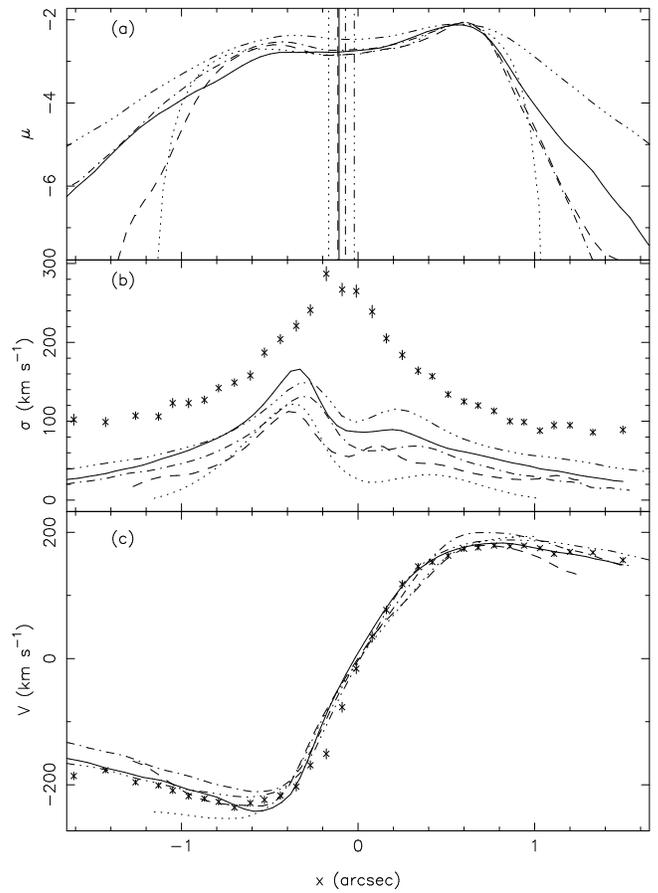}}
\caption{Comparison of the KB99 data with some of our projected models.  The observed kinematics along a cut passing through P1 and P2 are shown by crosses with error bars in panels (b) and (c) only; P1 is to the right of the origin and KB99 estimate that the BH (P2) is $0.080\pm0.011^{\prime\prime}$ to the left of the rotation center.  Our best model is shown by the full-drawn lines and is described in the text.  The disk masses in the other models are $M_d/M_\bullet = 0.001$ (dotted), 0.05 (dashed), 0.1 (dot-dashed) and 0.3 (triple-dot-dashed).  The radial extents of the disks and the imposed perturbations were all different and will be described in a later paper.}
\label{fig:fit}
\end{figure}

\section{Comparison with M31}
The brightest point of M31 is displaced slightly from the center of the bulge isophotes (Light \etal\ 1974) and the nucleus is now known to be double (Lauer \etal\ 1993), with two brightness peaks separated by $\sim 0.5^{\prime\prime}$.  The brighter peak is designated P1, while the fainter (P2) lies within $0.05^{\prime\prime}$ of the center of the outer nuclear isophotes and the bulge photocenter.  Ground-based integral-field spectroscopy by Bacon \etal\ (1994) revealed that P2 is close to the rotational center of the system, and that the dispersion peak coincides with neither brightness peak, and is displaced from P2 in a direction almost exactly opposite to P1.  Kormendy \& Bender (1999, hereafter KB99) confirm these features from spectroscopy with a single-slit with higher-resolution.  Bacon \etal\ (2000) report yet more detailed observations which confirm and extend previous work and which we discuss in a later paper.

In order to account for the observations, Tremaine (1995) proposed a model in which an eccentric disk of stars orbits a black hole (BH).  The fainter peak, P2, is presumed to be the location of the BH at the center of the large spheroidal bulge, while the brighter peak is caused by the lingering of stars at apocenter in the eccentric disk, viewed almost edge-on.  His model, which was designed to fit the photometric asymmetries only, placed stars in the disk on nested confocal elliptical orbits around the BH which dominates the total mass of the nucleus.  KB99 claim general agreement between their data and the kinematic predictions of Tremaine's model in both the shape and asymmetry of the circular speed profile about P2.

We here attempt a quantitative comparison between our self-consistent dynamical models and the KB99 data.  The symmetry in the bulge rotation curve well outside $3^{\prime\prime}$ allows KB99 to determine the systemic velocity of M31 to a precision of better than 1 km~s$^{-1}$.  They adopt a slightly different photometric bulge model from that proposed by Tremaine, but in either case the circular speeds arising from the central attraction of the bulge for M/L$_V = 4$ are negligible in comparison with those from the BH and nuclear disk in the inner $2^{\prime\prime}$.  KB99 subtract a scaled spectrum of the outer bulge from their spectra in the nuclear region to yield ``bulge subtracted'' estimates of the nuclear disk kinematics.  These data indicate streaming speeds rising to $\sim 230\;$km s$^{-1}$ on the P1 side, but to only $\sim 180\;$km s$^{-1}$ on the other side [see Figure \ref{fig:fit}(c)].  The asymmetry in velocities persists out to $\gtsim 3^{\prime\prime}$ from the rotation center, which they find is very close to P2 ($0.080\pm0.011^{\prime\prime}$ towards P1).

Although Tremaine suggests that the eccentric nuclear disk is coplanar with the outer disk of M31 (\ie\ inclined at $77^\circ$ from face-on), we ``view'' our razor-thin disk models edge-on in order to avoid having to thicken our model in some artificial manner.  We rotate our models so that the azimuthal angle in the disk-plane between disk major axis and the plane of the sky is $21^\circ$, as deduced by Tremaine.  We then compute the line of sight velocity and dispersion from the projected particle speeds using a kernel (Silverman 1986) with a fixed width chosen to approximate the spatial resolution of the KB99 data ($\sigma_* = 0.27^{\prime\prime}$).  We fit each of our models to the observed streaming velocity profile to $\pm 1^{\prime\prime}$ only, fitting the projected scales in both distance and velocity, together with a possible position offset, to minimize $\chi^2$.  Note that we do not allow the systemic velocity to be a free parameter in the fits.

A characteristic signature of the lopsided disk is the asymmetry in the streaming velocity on the two sides of the nucleus.  We have therefore focused on trying to fit the streaming velocity pattern.  Our most successful model is shown by the full-drawn curve in Figure \ref{fig:fit}(c).  This model has $M_d/M_\bullet = 0.2$, $a/R_0 = 2/3$ and both the inner and outer edges were initially elliptic with all orbits passing through the point $(-2R_0,0)$.  The evolved model (full-drawn curve) reproduces the $\sim 50\;$km~s$^{-1}$ difference between the peak velocities observed on the two sides tolerably well, and the streaming velocities in the model decrease from the central peaks roughly as do the data.  The full-drawn curves in panels (a) and (b) of Figure \ref{fig:fit} show, respectively, the projected mass density and velocity dispersion of the same model.  The mass density peaks at about $0.6^{\prime\prime}$ from the center, slightly farther than the observed position of P1 ($0.5^{\prime\prime}$); the peak value is $\sim 0.6\;$ mag above that at P2 which is nicely consistent with the cut through the HST photometry shown by KB99.  Also the position of the BH, marked by the vertical line, shows a small shift from the rotation center towards P2 of about the amount estimated by KB99.  The velocity dispersion profile does not fit the KB99 data at all well, however; the model is too cool everywhere and the position of the peak, while on the correct side, is too far from the center.

We have tried a large number of other models, but none fits the data as well.  Some of the better ones are shown by the broken curves in Figure \ref{fig:fit}.  A disk of virtually massless particles (dotted curves) with a broad radial extent has too great an asymmetry and the projected velocity dispersion is the lowest of all.  We have found that much heavier disks (\eg\ $M_d/M_\bullet = 0.3$, the triple-dot-dashed curve) tend not to show strong enough asymmetry in the streaming speed and the velocity dispersion is actually lower in this case than in our preferred model.

The scaling of our preferred model to the KB99 data require $R_0=0.523^{\prime\prime} = 1.95\;$pc (for an adopted distance of 770 kpc) and a velocity unit $= 301.6\;$km~s$^{-1}$.  With this scaling, one dynamical time is approximately $6\,300\;$years and the implied BH mass is $M_\bullet \sim 4.0 \times 10^7\;\hbox{M}_\odot$.

\section{Conclusions}
We have shown that lopsided modes in an almost-Keplerian annular disk can survive for at least 700 dynamical times, and probably indefinitely.  The modes, which are easily excited with a range of amplitudes in disks of differing radial extents, survive with significant amplitude to disk masses at least as large as 30\% of the central mass.  If the modes are also present in collisional, or gaseous disks, they could be of relevance to the formation of massive planets or small companion stars from a disk surrounding a primary, or for the persistence of asymmetric accretion disks in other contexts.

The existence of these long-lived modes offers considerable support to Tremaine's (1995) eccentric disk model for the double nucleus of M31, but we have been unable to create a model that matches all aspects of the observations.  One of our models reproduces some significant kinematic and photometric features of the M31 nucleus quite well, including the off-center brightness peak, the asymmetry in the streaming velocities, and approximately correct location of the velocity zero between P1 and P2.  The velocity dispersion peak is also off-center, but not in quite the right place and the model is altogether too cool.  Tremaine has suggested to us that two-body relaxation is important in the nucleus, a process largely absent from our models.  It remains to be seen whether a model with the considerably greater random motion needed to fit the observed dispersion profile could continue to fit the streaming and photometric profiles.

\acknowledgments
We thank Scott Tremaine for his interest and advice and the referee for a thoughtful report.  This work was supported by NASA LTSA grant NAG 5-6037.

\end{document}